\documentclass[12pt]{article}

\catcode`\@=11
\@addtoreset{equation}{section}

\global\arraycolsep=2pt
\usepackage{amsbsy,amssymb,latexsym,amsfonts,amsmath}
\usepackage{graphicx,color}

\allowdisplaybreaks

\begin{document}
\begin{flushright}
\parbox{4.2cm}
{}
\end{flushright}

\vspace*{0.7cm}

\begin{center}
{\Large \bf 
CP-violating CFT and trace anomaly}
\vspace*{1.5cm}\\
{Yu Nakayama}
\end{center}
\vspace*{1.0cm}
\begin{center}
{\it Institute for the Physics and Mathematics of the Universe,  \\ Todai Institutes for Advanced Study,
University of Tokyo, \\ 
5-1-5 Kashiwanoha, Kashiwa, Chiba 277-8583, Japan}
\vspace{3.8cm}
\end{center}

\begin{abstract} 
It is logically possible that the trace anomaly in four dimension includes the Hirzebruch-Pontryagin density in CP violating theories. 
  Although the term vanishes at free conformal fixed points, we realize such a possibility in the holographic renormalization group and show that it is indeed possible. The Hirzebruch-Pontryagin term in the trace anomaly may serve as a barometer to understand how much CP is violated in conformal field theories.
\end{abstract}

\thispagestyle{empty} 

\setcounter{page}{0}

\newpage

\section{Introduction} 
The very existence of our human being relies on CP violation in an essential manner through Sakharov's condition on the baryogenesis. While the breaking mechanism of CP symmetry is encoded in the standard model by three generations and phases in Yukawa couplings as well as Yang-Milles theta terms, it is a difficult question to measure how much CP is violated in a given quantum field theory in a qualitative manner. The ultimate goal of our study is to seek the possibility to use the CP violating contribution to the trace anomaly for such a candidate.

Surprisingly, CP violation in conformal field theories have been rarely studied in literatures. To some extent, it is due to the restricted viewpoint or more or less folklore that the conformal symmetry is obtained by adding ``inversion" symmetry to the Poincar\'e group. Although it is true that the successive application of inversion, translation and the second inversion gives the special conformal transformation, the converse is not necessarily true: the conformal field theory may not be invariant under the inversion. Mathematically speaking, the conformal group is $SO(2,d)$, but with the inversion it is enhanced to $O(2,d)$. Clearly, the latter contains the former, but it is the former $SO(2,d)$ that is only required for the symmetry of quantum field theories.

The inversion, in the radial quantization, is nothing but the time reversal. The CPT-theorem tells us that whenever CP is violated, the time reversal must be broken. Therefore, we conclude that the CP violating conformal field theory cannot possess invariance under inversion. Technically speaking, studying the constraint from  invariance under inversion is much convenient and probably the easiest way to obtain the form of correlation functions that are invariant under the special conformal transformation. The argument here, however suggests that the imposition might be too restrictive, and indeed it was demonstrated that it is the case. 

In this paper, we discuss these usually overlooked aspects of conformal field theories with CP-violation. We discuss possible CP-violating terms in correlation functions of the energy-momentum tensor and its trace anomaly. In particular, we argue that it is entirely legitimate for the trace anomaly to have the Hirzebruch-Pontryagin density, which is parity odd. 
It is known that CP is preserved in any unitary free conformal field theories in four space-time dimension, so it is very difficult to show how they could appear in actual computations. In this paper, we take an alternative route by studying strongly coupled dual field theories with the help of the holographic renormalization group method. We show its actual existence by studying the holographic renormalization of the bulk CP violating gravitational theory.

The organization of the paper is as follows. In section 2, we discuss the consequence of CP-violation in correlation functions of the energy-momentum tensor and the trace anomaly. In particular, we argue that the Hirzebruch-Pontryagin density is a legitimate candidate of the trace anomaly in four-dimension. In section 3, we show how the CP-violating Hirzebruch-Pontryagin density can appear in the trace anomaly from the holographic renormalization method in five-dimensional AdS space with a CP violating  action. In section 4, we present implications of our result in relation to the recently proved a-theorem for conformal fixed points and conclude.

\section{CP-violation in energy-momentum tensor}
Conformal algebra is the only natural space-time extension of the Poincar\'e algebra 
\begin{align}
i[J^{\mu\nu},J^{\rho\sigma}] &=
\eta^{\nu\rho}J^{\mu\sigma}-\eta^{\mu\rho}J^{\nu\sigma} -
\eta^{\sigma\mu}J^{\rho\nu} + \eta^{\sigma\nu}J^{\rho\mu} \cr
i[P^\mu,J^{\rho\sigma}] &= \eta^{\mu\rho}P^{\sigma} -
\eta^{\mu\sigma} P^\rho \cr[P^\mu,P^\nu] &= 0 \  \label{alg1}
\end{align}
with the dilatation symmetry
\begin{align}
[P^\mu,D] &= i P^\mu \cr [J^{\mu\nu},D] &= 0  \ . \label{alg2}
\end{align}
It possesses the additional generators $K^\mu$ called special conformal transformation:
\begin{align}
[K^\mu,D] &= -iK^\mu \cr [P^\mu,K^\nu] &= 2i\eta^{\mu\nu}D+
2iJ^{\mu\nu} \cr [K^\mu,K^\nu] &= 0 \cr [J^{\rho\sigma},K^\mu] &=
i\eta^{\mu\rho} K^\sigma - i\eta^{\mu\sigma} K^\rho \ . \label{alg3}
\end{align}
It is isomorphic to the Lie algebra $SO(2,d)$, and it is known that the conformal {\it algebra}, consisting of  \eqref{alg1}, \eqref{alg2} and \eqref{alg3}, is the largest bosonic space-time symmetry that acts non-trivially on S-matrix (in $d>2$) of massless particles \cite{Haag:1974qh}.\footnote{Note that the algebraic consideration in \cite{Haag:1974qh} did not say anything about the global structure of the group. The inclusion of the fermion necessitates the inclusion of the spinor representations, and the symmetry {\it group} must be its double cover. Additional discrete space-time parities (e.g. CP symmetry) may or may not be the symmetry of the full theory.}

For later purposes, let us review the salient feature of the energy-momentum tensor. The Poincar\'e invariance \eqref{alg1} demands that it is symmetric and conserved: $\partial^\mu T_{\mu\nu} = \partial^\mu T_{\nu\mu} = 0$. The dilatation invariance \eqref{alg2} demands that its trace is given by the divergence of the so-called Virial current \cite{Wess}\cite{Mack}\cite{Coleman:1970je}: $T^{\mu}_{\ \mu} = \partial^\mu J_\mu$. Finally, the conformal invariance \eqref{alg3} demands that the Virial current can be expressed as a derivative: $J_\mu = \partial^\nu L_{\mu\nu}$ so that the energy-momentum tensor can be improved to be traceless (see e.g. \cite{Polchinski:1987dy}).

The alternative but restrictive way to see the conformal group would be to consider the inversion $x^\mu \to -\frac{x^\mu}{x^2}$. The successive transformation of inversion, translation and the second inversion gives the special conformal transformation $x^\mu \to \frac{x^\mu -a^\mu x^2}{1-2a^\mu x_\mu + a^2 x^2}$. Obviously the theory which is invariant under the inversion (together with the Poincar\'e invariance) is invariant under the full conformal transformation, but the inversion is not necessarily required to have $SO(2,d)$ conformal symmetry. Indeed, with the inversion, the symmetry {\it group} (rather than the algebra) is enhanced to  $O(2,d)$ (see e.g. \cite{Weinberg:2010fx}), and the inversion lies in the disconnected component of the group.

In many correlation functions, however, it happens that the effect of the CP violation cannot appear simply due to the strong constraint coming from the special conformal transformation. For instance, all the scalar correlation functions are insensitive to the CP violation. To see a possibility to have a non-trivial CP violating contribution, let us consider the two-point function of the  energy-momentum tensor on the flat space-time \cite{Osborn:1993cr}\cite{Erdmenger:1996yc}:
\begin{align}
\langle T_{\mu\nu}(x) T_{\alpha\beta}(0) \rangle_{\text{CP even}}  = c\frac{ \mathcal{I}_{\mu\nu;\alpha\beta}(x)}{x^{2d}} \ ,
\end{align}
where $\mathcal{I}_{\mu\nu;\alpha\beta} = \frac{1}{2}(I_{\mu\alpha}I_{\nu \beta} + I_{\mu\beta} I_{\nu\alpha}) - \frac{1}{d}\delta_{\mu\nu} \delta_{\alpha\beta}$ with $I_{\mu\nu} = \delta_{\mu\nu} -2\frac{x_\mu x_\nu}{x^2}$ is the fixed parity even tensor, and the whole structure is entirely dictated by the $SO(2,d)$ invariance with or without inversion.
This result is valid in any space-time dimension except for $d=3$, and we see that there is no CP violating term here. The exception appears only in $d=3$. There, we may have a possible parity violating term (see e.g. \cite{Leigh:2003ez}\cite{Maldacena:2011nz}):\footnote{In this paper, the Levi-Civita symbol is defined as a tensor rather than a tensor density.}
\begin{align}
\langle T_{\mu\nu}(x) T_{\alpha\beta}(0) \rangle_{\text{CP odd}}  = w (\epsilon_{\mu \alpha \sigma}(\partial_\nu \partial_\beta - \eta_{\nu \beta} \partial^2) \partial^\sigma \delta^4(x) + \text{sym}) \ , \label{emt}
\end{align}
where sym means the symmetrization under $\mu \leftrightarrow \nu$ and $\alpha \leftrightarrow \beta$.
This CP violating term is classified as a contact term (due to the delta-function in \eqref{emt}), and the physical significance is a little bit subtle. The appearance of the CP violating term in three-dimension has a deep connection with the parity anomaly because integrating out an odd number of massive fermions gives rise to effective gravitational Chern-Simons action that would generate the contact term \eqref{emt}.  How this parity violating contact term can arise in the holographic computation was pursued in some literatures (e.g. \cite{Leigh:2003ez}\cite{Leigh:2008tt}). It is ultimately due to the gravitational theta term \cite{Deser:1980kc} $\int d^4x \sqrt{g} \theta_{G} \epsilon^{\alpha\beta \gamma \delta} R_{\alpha\beta \mu\nu}R^{\mu\nu}_{\ \ \gamma \delta}$ that violates the CP symmetry in the bulk four-dimensional gravitational theory.\footnote{The bulk gravitational theta term reduces to the boundary gravitational Chern-Simons term, and it directly gives the  parity violating contact term in the boundary correlation function.} 

A somewhat related manifestation of the CP violation in (conformal) field theories in three space-time dimensional space-time is the expectation value of the energy-momentum tensor in the curved background. It can possess the parity violating term \cite{deHaro:2008gp}:
\begin{align}
\left \langle T_{\mu\nu} \right \rangle_{\text{CP odd}} = \tilde{w} C_{\mu\nu}
\end{align}
where $C_{\mu\nu} = \epsilon_{\mu \sigma \rho} D^{\sigma} (R^{\rho}_{\ \nu} -\frac{1}{4} R \delta^{\rho}_{\ \nu}) $ is the (traceless and conserved) Cotton tensor, which is intrinsic to three-dimension. Again the explicit appearance of $\epsilon_{\mu \sigma \rho}$ suggests violation of parity. Since the Cotton tensor is traceless, there is no trace anomaly here.

The main focus of this paper is in four-dimensional space-time. As we have seen, in four dimension, there is no CP violating term in two-point function of the energy-momentum tensor in conformal field theories. This is even true if we relax the conformal invariance to the mere scale invariance (with the Poincar\'e invariance intact) \cite{Dorigoni:2009ra}. 

In four-dimension, the dimensional analysis demands that the most general possibility of the trace anomaly (for an earlier review, see \cite{Duff:1993wm}\cite{Deser:1996na} and references therein) be
\begin{align}
T^{\mu}_{\ \mu}  = cF + a G  + b R^2 + b' \Box R + e \epsilon^{\alpha\beta \gamma \delta} R_{\alpha\beta \mu\nu}R^{\mu\nu}_{\ \ \gamma \delta} \  , 
\end{align}
aside from  possible operator violation in non-conformal field theories. Here $F = R^{\alpha\beta\gamma\delta} R_{\alpha\beta\gamma \delta} - 2R^{\alpha\beta} R_{\alpha \beta} + \frac{1}{3}R^2$ is the square of the Weyl tensor while $G= R^{\alpha\beta\gamma\delta} R_{\alpha\beta\gamma \delta} - 4R^{\alpha\beta} R_{\alpha \beta} + R^2$ is the Euler density. The last term $\epsilon^{\alpha\beta \gamma \delta} R_{\alpha\beta \mu\nu}R^{\mu\nu}_{ \ \ \gamma \delta}$ is the parity odd Hirzebruch-Pontryagin density.

In conformal field theories, the Wess-Zumino consistency condition, which is essentially the requirement that the successive applications of the Weyl transformation must commute because the Weyl transformation is abelian, demands that $b=0$ \cite{Bonora:1983ff}\cite{Bonora:1985cq}\cite{Cappelli:1988vw}\cite{Osborn:1991gm}. In addition, $b'$ term is trivial in the sense that we can always remove it by adding the local counterterm $\int d^4x \sqrt{g} R^2$. See \cite{Deser:1976yx}\cite{Deser:1993yx} for further discussions on the classification of the CP non-violating terms.

The parity odd term, which has been neglected in many literatures, cannot be logically excluded. It satisfies the Wess-Zumino consistency condition \cite{Bonora:1985cq} because the Hirzebruch-Pontryagin density is Weyl invariant \cite{Deser:1996na}. It may serve as a barometer that measures the violation of the CP in a given conformal field theory. In the rest of the paper, we would like to  investigate the structure and the consequence of this term in the trace anomaly. 

Of course, the Wess-Zumino consistency condition does not tell us whether we indeed have such a term. For this purpose, we need an explicit computation. What bothers us here is that the simplest computation of the trace anomaly is done in free field theories, necessarily at one-loop. It turns out that all the free conformal invariant field theories in four-dimension are actually invariant under CP, so the free field computation always predicts $e=0$. 

If this were the chiral anomaly, where the one-loop exactness is proved (known as the Adler-Bardeen theorem), this would be the end of the pursuit. If there were no anomaly at one-loop, there would be none in the full computation (at least to all orders in perturbation theory).
Fortunately, the one-loop exactness is not true for the trace anomaly. We already know that the central charge $a$ and $c$ are not one-loop exact, and there is no reason why it should be so for the CP odd term with the ``CP violating central charge" $e$. Indeed, to pick up the CP odd term in the trace anomaly, we need at least two-loop or higher (or even non-perturbative) contributions.\footnote{A non-perturbative contribution is certainly important when the CP violation is due to the Yang-Mills theta term.}

To counter the suspicion for the very possibility to have the Hirzebruch-Pontryagin density in the trace anomaly, we would like to mention that there exists a free field computation for the self-dual (or anti-self-dual) two-form gauge field (i.e. $B_{\mu\nu} = \pm \epsilon_{\mu\nu\rho\sigma}B^{\rho\sigma}$) in four-dimensional Euclidean-signatured space. The explicit heat kernel analysis of the propagator showed that the self-dual two-form gauge field gives rise to the trace anomaly \cite{Duff:1980qv}\footnote{In the formula here, we have not introduced the ghost contribution which in any case would not affect the Hirzebruch-Pontryagin density.}
\begin{align}
 T^{\mu}_{\ \mu} = \frac{1}{180(4\pi)^2} \left( 33R^{\alpha\beta\gamma\delta} R_{\alpha\beta\gamma \delta} - 93 2R^{\alpha\beta} R_{\alpha \beta} + \frac{45}{2}R^2 - 12 \Box R +  30  \epsilon^{\alpha\beta \gamma \delta} R_{\alpha\beta \mu\nu}R^{\mu\nu}_{\ \ \gamma \delta} \right) 
\end{align}
while the anti-self dual two-form gives rise to the trace anomaly
\begin{align}
 T^{\mu}_{\ \mu} = \frac{1}{180(4\pi)^2} \left( 33R^{\alpha\beta\gamma\delta} R_{\alpha\beta\gamma \delta} - 93 2R^{\alpha\beta} R_{\alpha \beta} + \frac{45}{2}R^2 - 12 \Box R -  30  \epsilon^{\alpha\beta \gamma \delta} R_{\alpha\beta \mu\nu}R^{\mu\nu}_{\ \  \gamma \delta} \right) 
\end{align}
The geometrical reason why we obtained the Hirzebruch-Pontryagin density is rather clear: the integrated anomaly will give the Hirzebruch signature, and it directly relates the zero-mode of the two-form gauge field through the index theorem. The heat kernel computation is a simple manifestation of the famous Hirzebruch signature theorem: $n(B^+) - n(B^-) = \frac{1}{48\pi^2} \int d^4x \sqrt{g}  \epsilon^{\alpha\beta \gamma \delta} R_{\alpha\beta \mu\nu}R^{\mu\nu}_{\ \ \gamma \delta}$ 

The drawback of the free field computation here is that the restriction to the (imaginary) self-dual (or anti-self-dual) two-form field in Minkowski signature leads to non-unitarity. Therefore, the self-dual (or anti-self-dual) two-form gauge field considered here is not physical. However, it clearly demonstrates that the possibility of the Hirzebruch-Pontryagin density in the trace anomaly is not something that can be thrown away immediately.\footnote{Indeed, the appearance of the Hirzebruch-Pontryagin density is ubiquitous in non-unitary free field computations. It has been demonstrated \cite{Dowker:1976zf}\cite{Christensen:1978gi}\cite{Christensen:1978md} that whenever the Lorentz group representation is not  real (or not symmetric under the exchange of the two $SU(2)$ in Euclidean signature),  the contribution is non-zero. We would like to thank M.~Duff for the correspondence.} In the next section, we try to realize the emergence of the Hirzebruch-Pontryagin density in the trace anomaly from the holographic renormalization group approach.

The appearance of the CP violating Hirzebruch-Pontryagin density in the trace anomaly will affect the structure of the three-point functions of the energy-momentum tensor in flat four-dimensional space-time. In generic $d$-dimensional space-time, the structure of the three-point function of the energy-momentum tensor in conformal field theories with no parity violation, was studied in \cite{Osborn:1993cr}\cite{Erdmenger:1996yc}. Due to its complexity in appearance, we will not show the whole structure here, but if we take trace of one of the energy-momentum tensor, the result simplifies a bit and it reduces to contact terms
\begin{align}
& \langle T^{\mu}_{\ \mu} (x) T_{\sigma\rho}(y)T_{\alpha\beta}(z) \rangle_{\text{CP even}} \cr 
&= 2 (\delta^4(x-y) + \delta^4(x-z)) \langle T_{\sigma \rho}(y) T_{\alpha\beta}(z) \rangle - 4(c \mathcal{A}^F_{\sigma \rho, \alpha \beta} + a \mathcal{A}^G_{\sigma \rho, \alpha \beta} ) \ ,
\end{align}
where
\begin{align}
\mathcal{A}^F_{\sigma \rho, \alpha \beta} &=-8 \mathcal{E}^C_{\sigma \kappa \lambda \rho, \alpha \gamma \delta \beta} \partial^\kappa \partial^\lambda \delta^4(x-y) \partial^\gamma \partial^\delta \delta^4(x-z) \cr
\mathcal{A}^G_{\sigma \rho, \alpha \beta} &= \epsilon_{\sigma \alpha \gamma \kappa} \epsilon_{\rho \beta \delta \lambda} \partial^\kappa \partial^\lambda(\partial^\gamma \delta^4 (x-y)\partial^\delta \delta^4(y-z)) + \text{sym}  \ .
\end{align}
Here  $ \mathcal{E}^C_{\sigma \kappa \lambda \rho, \alpha \gamma \delta \beta}  = \partial C_{\mu\sigma\rho\nu}/ \partial C^{\alpha \gamma \delta \beta}$, and we can find the explicit form in Appendix A of \cite{Erdmenger:1996yc}.

 Now with the CP violation, the three-point function must possess the additional term
\begin{align}
\langle T^{\mu}_{\ \mu}(x) T_{\sigma \rho}(y) T_{\alpha \beta}(z) \rangle_{\text{CP odd}} = e \epsilon_{\sigma \alpha \epsilon \kappa} (\partial_\beta \partial_\rho - \partial^2 \delta_{\beta \rho})[ \partial^\epsilon \delta^4(x-y) \partial^{\kappa} \delta^4(x-z) ] + (\text{sym}) \ .  \label{mmv}
\end{align}
We have not studied the structure of the full three-point functions with the CP violation due to its complexity. It would be interesting to see its structure, and verify whether there is any other structure whose origin is not related to \eqref{mmv}. In particular, it would be exciting to see whether we have any free additional parameters besides the ``CP violating central charge" $e$ to completely determine the three-point function of the energy-momentum tensor with  CP violation.

\section{Holographic realization}
We have seen that the free field (or one-loop) computation of the trace anomaly  does not lead to the CP violating Hirzebruch-Pontryagin density because all the free unitary conformal field theories preserve CP. We may attempt computing the higher loop corrections, but in this section, we take an alternative approach based on the holographic renormalization group \cite{Henningson:1998gx} to purse its possibility in strongly coupled dual field theories.

The toy model we will consider is the generalization of the model studied in \cite{Nakayama:2009qu}\cite{Nakayama:2009fe}\cite{Nakayama:2010ye} for a gravity dual of scale invariant but not conformal field theories. It is 
given by the five-dimensional  Einstein gravity coupled with a self-interacting vector field with the action
\begin{align}
S_{\mathrm{bulk}}  = \int d^5x \sqrt{g}\left[ \frac{1}{2\kappa_5}(R - 2\Lambda) + \left(\frac{1}{4}F^2 + V(A_M A^M) \right) \right]\ .
\end{align}
In order to break the CP of the bulk gravity, we introduce the gravity-vector Chern-Simons-like term
\begin{align}
S_{\mathrm{CSL}} = q\int d^5x \sqrt{g} \epsilon^{LMNPQ} A_L R_{MN IJ} R_{\ \ PQ}^{IJ} \label{CSL}
\end{align}
To make the variation principle well-defined, we may want to introduce the boundary term \cite{Landsteiner:2011iq}
\begin{align}
S_{\mathrm{CSK}} = -8q\int_{\partial M} d^4x\sqrt{h} n_M \epsilon^{MNPQR}A_N K_{PL} D_QK^L_R \ , 
\end{align}
where $n_M$ is the normal vector and $K_{AB}$ is the extrinsic curvature.

The Chern-Simons-like term \eqref{CSL} is imperative to break the parity invariance of the gravitational bulk theory, and it is the same action that would generate the gravitational chiral anomaly of a conserved current in the boundary theory if $A_{M}$ were a gauge field. This fact will be crucial in the field theory interpretation we will discuss later.

The condition of the scale invariance dictates that the metric must take the form of $AdS_5$
\begin{align}
ds^2 = g_{MN}dx^M dx^N =  R^2_{AdS_5}\frac{dz^2 + \eta_{\mu\nu}dx^\mu dx^{\nu}}{z^2} \ . 
\end{align}
We choose the potential $V(A_M A^M) = \sum_n a_n (A_M A^M)^n$ so that $A = A_{M} dx^M = a\frac{dz}{z}$ is the solution of the equations of motion. The potential $V(A_M A^M)$ explicitly breaks the gauge invariance of the vector field $A_M$. Alternatively one can regard $A_M$ as the gauge fixed version of the Stueckelberg field with the higher derivative covariant action $ \sum_n a_n (\partial_M \phi - A_M)^{2n} $ for the  Stueckelberg scalar $\phi$ in the unitary gauge $\phi=0$.   

We can see that the vector condensation does not backreact to the metric, so the geometry is still AdS space. 
Indeed, the solution is the same one studied in \cite{Nakayama:2009qu}\cite{Nakayama:2009fe}\cite{Nakayama:2010ye} in the context of the gravity dual of scale invariant but non-conformal field theory. Clearly, the extra Chern-Simons-like term did not affect the classical solution because $dA = 0$.

We want to study the holographic renormalization of the system by considering the Fefferman-Graham expansion of the metric 
\begin{align}
\frac{ds^2}{R^2_{AdS_5}} = \frac{dz^2}{z^2} + \frac{h_{\mu\nu} dx^{\mu} dx^{\nu}}{z^2}
\end{align}
with 
\begin{align}
h_{\mu\nu} = h_{\mu\nu}^{(0)} + z^2 h_{\mu\nu}^{(2)} + z^4h_{\mu\nu}^{(4)} + \cdots  
\end{align}
and evaluating the on-shell action. The on-shell action is divergent so we introduce the cutoff at $z= \epsilon$. The logarithmic dependence of the on-shell action on the cutoff is then interpreted as the holographic trace anomaly.
Aside from the usual term that gives the holographic realization of the parity preserving trace anomaly $T^{\mu}_{\ \mu} = cF + a G$, we can immediately find the counter-term necessary from the Chern-Simons-like term 
\begin{align}
S_{\mathrm{CSL}}^{(0)} = qa\log \epsilon \int_{\partial M} d^4x \sqrt{h}  \epsilon^{\alpha\beta \gamma \delta} R_{\alpha\beta \mu\nu}R^{\ \ \mu\nu}_{\gamma \delta}  \label{onsa}
\end{align}
by noting $ A = a\frac{dz}{z} = a d(\log z)$ is exact and by using the Stokes theorem. This leads to the additional CP violating contribution to the trace anomaly of the dual boundary field theory
\begin{align}
T^{\mu}_{\ \mu}|_{\text{CP odd}}  = e \epsilon^{\alpha\beta \gamma \delta} R_{\alpha\beta \mu\nu}R^{\ \ \mu\nu}_{\gamma \delta} \   
\end{align}
with $e=qa$.
In this way, we have demonstrated how the CP violating Hirzebruch-Pontryagin density can arise in the trace anomaly from the holographic computation.

We would like to give the interpretation of the CP-violating contribution to the trace anomaly in this model from the dual field theory. First of all, we recall that in scale but non-conformal field theory, the trace of the energy-momentum tensor is non-zero even in the flat space-time: rather it is given by the divergence of the so called Virial current
\begin{align}
 T^{\mu}_{\ \mu} = \partial^\mu J_\mu \ . \label{virir}
\end{align}
In our holographic description, the vector condensation $A = a\frac{dz}{z}$ is dual to the existence of the non-zero Virial current \cite{Nakayama:2010wx}. As we have mentioned, the current model serves as the gravity dual of sale invariant but non-conformal field theory due to the existence of the non-zero Virial current.

In our CP violating scenario, we assume that the Virial current contains the CP violating term, or in other words it is a chiral current. While the Virial current is not conserved, it is typical that some sort of equations of motion were used in deriving the equality \eqref{virir}.\footnote{In renormalizable field theories, the trace of the energy-momentum tensor is typically given by $\phi^4$ terms or Yukawa terms. To connect them to divergence of currents of dimension 3, we use the equations of motion.} Now, the key idea is that once we evaluate the equality \eqref{virir} in the curved background, it is expected that the gravitational chiral anomaly for the Virial current $J_\mu$ would give an additional piece in \eqref{virir}. Since we know that the gravitational chiral anomaly must contain the Hirzebruch-Pontryagin term \cite{Eguchi:1976db} as $D^\mu J_\mu = \kappa \epsilon^{\alpha\beta \gamma \delta} R_{\alpha\beta \mu\nu}R^{\ \ \mu\nu}_{\gamma \delta}  + (\text{non-anomalous term})$, we expect
\begin{align}
 T^{\mu}_{\ \mu} = D^\mu J_{\mu} - \kappa \epsilon^{\alpha\beta \gamma \delta} R_{\alpha\beta \mu\nu}R^{\ \ \mu\nu}_{\gamma \delta}  + \cdots
\end{align}
with possible CP non-violating contribution to the trace anomaly. 

This is exactly what was happening in the holographic computation. The Chern-Simons-like term we added is nothing but declaring that the current under consideration has a gravitational chiral anomaly. The fact that it is not gauge field and has a vacuum expectation value $a \frac{dz}{z}$ is the manifestation that it is not conserved and it is rather the Virial current appearing in the trace of the energy-momentum tensor. The combination of these two led to the CP violating Hirzebruch-Pontryagin term in the trace anomaly as expected in the field theory argument above.

Finally, let us note the following fact. The Virial current has an ambiguity so that it is only defined up to a conserved current. When the theory possesses an  extra conserved chiral current $J_\mu^c$ with gravitational chiral anomaly (i.e. $D^\mu J_\mu^c = \kappa_c\epsilon^{\alpha\beta \gamma \delta} R_{\alpha\beta \mu\nu}R^{\ \ \mu\nu}_{\gamma \delta}$), we can augment the Virial current with that conserved chiral current so that the new ``improved" Virial current $\tilde{J}_\mu = J^\mu - \frac{\kappa}{\kappa_c} J_\mu^c$ shows no parity odd term after taking the divergence: $T^{\mu}_{\ \mu} = D^\mu \tilde{J}_{\mu} $. 

The same thing can be done in the holographic computation. The conserved chiral current with the gravitational chiral anomaly can be implemented as a bulk gauge field $A^c$ with the gauge-gravity Chern-Simons interaction $q_c\int d^5x \sqrt{g} \epsilon^{LMNPQ} A_L^c R_{MN IJ} R_{\ \ PQ}^{IJ}$. Now, we do the (large) gauge transformation $A^c = -\frac{qa}{q_c} \frac{dz}{z}$ and evaluate the on-shell action by using the Stokes theorem. It again shows the logarithmic divergence with the Hirzebruch-Pontryagin density. Then we can cancel the logarithmic divergence of the on-shell action coming from the Virial current \eqref{onsa} with this new contribution from the conserved current.

On one hand, this illustrates how the definition of the trace of the energy-momentum tensor can be ambiguous with more conserved currents in scale invariant but non-conformal field theories, but on the other hand, it demonstrates clearly that there is nothing wrong with having the Hirzebruch-Pontryagin density in the trace anomaly in conformal field theories. We may just discard the contribution from the genuine Virial current in the above discussions, and see the CP violating terms appear.
It would be interesting to see how much that ambiguity can be fixed in conformal field theories from the purely field theory argument.\footnote{It is important to note that it is not mandatory to cancel the CP violating term even if we have such an option. The choice of the large gauge parameter determines the theory on the curved space-time and our statement is that simply there are as many choices. The situation is closer to the parity violating contact term in three-dimensional conformal field theory reviewed in section 2. The choice of the contact term defines different theories (see \cite{Witten:2003ya} for a related discussion).} For instance, when the theory does not possess any conserved chiral current with gravitational anomaly, there is no possibility of such. 

The holographic model discussed in this section is based on the scale invariant but non-conformal field theory, and we do not know any cleaner, preferably conformal, holographic theories that show the CP violating trace anomaly except for the  possibility to use the large gauge transformation mentioned in the last paragraph. Since it may be possible that there is no unitary scale invariant but non-conformal field theory in four-dimension \cite{Polchinski:1987dy}\cite{Dorigoni:2009ra}\cite{Nakayama:2010wx}, there may be a hidden no-go theorem to have the Hirzebruch-Pontryagin density in unitary conformal field theories.
We would like to leave this field theoretical question for future studies.

\section{Discussions}
In this paper, we have studied how the Hirzebruch-Pontryagin density can appear in the trace anomaly when the theory under consideration breaks CP symmetry. We have demonstrated its possibility in the holographic renormalization computation. 
Although we did not discuss it in the main part of the paper, if we introduced the background gauge field for the global symmetry, we would also be able to introduce the CP violating Chern-Pontryagin density $\epsilon^{\mu\nu \alpha\beta} \hat{F}_{\mu\nu} \hat{F}_{\alpha\beta}$, where $\hat{F}_{\mu\nu}$ is the corresponding field strength, in the trace anomaly of CP violating conformal field theories. The gravity dual would require the vector-gauge Chern-Simons-like term $\int d^5x \sqrt{g} \epsilon^{MNLPQ} A_{M} \hat{F}_{NL}\hat{F}_{PQ}$. By using the same mechanism discussed in the previous section, we are able to reproduce the trace anomaly with the Chern-Pontryagin density.

In a similar manner, we may imagine that the chiral current anomaly could include the parity {\it even} term such as $\partial_\mu J^\mu_5 = \hat{e} F_{\mu\nu}F^{\mu\nu}$ in addition to the conventional parity odd term  $ \epsilon^{\mu\nu\alpha \beta} {F}_{\mu\nu} {F}_{\alpha\beta}$ in CP violating theories. After all, for the $U(1)$ symmetry, it is known that the Wess-Zumino consistency condition does not forbid it. We know, however, according to the Adler-Bardeen theorem, at least to all orders in perturbation theory, there cannot be such a contribution. Not surprisingly, we did not find any gravity computation that gives the corresponding result as far as we tried. It would be interesting to give a proof of the no-go theorem from the holographic viewpoint.

Recently, the ingenious proof of the a-theorem was demonstrated in \cite{Komargodski:2011vj}\cite{Komargodski:2011xv} when the flow is between two conformal field theories. The theorem states that a function called ``a", which is nothing but the coefficient in front of the Euler term in the trace anomaly, always satisfies the inequality $a_{\mathrm{UV}} > a_{\mathrm{IR}}$  along the renormalization group flow for any pairs of conformal field theories.
The crucial idea for its proof is to consider the  Wess-Zumino term associated with the trace anomaly in the spirit of the anomaly matching. There, they only discussed the CP conserving trace anomaly, so it is important to understand what happens if we allow the CP violating trace anomaly we have discussed in this paper.

To begin with, let us consider the conformal invariant (non-universal) CP violating action for dilaton $\tau$. There is none at the two-derivative level. At the four-derivative level, the only possible new term is the Hirzebruch-Pontryagin density: 
\begin{align}
S^{\mathrm{eff}}_{\text{CP odd}} =  \hat{e} \int d^4 x \sqrt{\hat{g}} \hat{\epsilon}^{\alpha\beta \gamma \delta} \hat{R}_{\alpha\beta \mu\nu}\hat{R}^{\mu\nu}_{\ \ \gamma \delta} \ ,
\end{align}
where $\hat{g}_{\mu\nu} = e^{-2\tau} g_{\mu\nu}$.
 If we evaluate this non-universal term in the flat space, it obviously vanishes since the Hirzebruch-Pontryagin density is a total derivative, so there is no additional CP violating four-derivative non-universal interaction for the dilaton.

Now, we consider the Wess-Zumino term. To cancel the ultraviolet CP violating trace anomaly, we have to introduce the Wess-Zumino term
\begin{align}
(e_{\mathrm{UV}}-e_{\mathrm{IR}}) \int d^4x \sqrt{g} \tau {\epsilon}^{\alpha\beta \gamma \delta} {R}_{\alpha\beta \mu\nu}{R}^{\mu\nu}_{\ \ \gamma \delta} \ .
\end{align}
Like c-anomaly and unlike a-anomaly, there is no further term necessary to complete the Wess-Zumino term because the Hirzebruch-Pontryagin density is Weyl invariant. Consequently, if we evaluate the Wess-Zumino term in the flat space-time, it vanishes and there is no CP violating contribution to the dilaton effective action. We thus conclude that the proof of a-theorem in \cite{Komargodski:2011vj}\cite{Komargodski:2011xv} is not affected by the existence of the CP violating contribution to the trace anomaly.

As we have mentioned in section 2, in non-conformal field theories, there may exist another ``central charge" $b$ that appears in $R^2$ term of the trace anomaly. In principle such a term can appear in the holographic computation of the scale invariant but non-conformal field theory. In the model we have studied in section 3, we did not find such a contribution but it is plausible that it would appear if we added the term like $R_{MN}A^MA^N$ to the gravity action, which breaks the AdS isometry in the gravity sector equation of motion with the vector condensation. It would be also interesting to verify how precisely $b$ term does or does not affect the derivation of the a-theorem for non-conformal but scale invariant fixed points (see \cite{Nakayama:2011wq} for related studies).\footnote{In  \cite{Nakayama:2011wq}, the scale invariant field theory is embedded in a conformal field theory to discuss the a-theorem for scale but non-conformal field theories, and the $b$ anomaly must have vanished for the consistency of the embedding. The treatment and the argument given there is self-consistent, but it remains open how the $b$ anomaly is cancelled in actual models. Furthermore, after all, we have not succeeded in deriving a-theorem in scale but non-conformal field theories, so the conformal embedding might not be a good idea, and then non-zero $b$ may play a role.}

\section*{Acknowledgements}
The author would like to thank Y.~Tachikawa for discussions. The work is supported by the World Premier International Research Center Initiative of MEXT of
Japan. 

\appendix

\end{document}